\documentclass[twocolumn,showpacs,preprintnumbers,amsmath,amssymb]{revtex4}
\usepackage{graphicx}% Include figure files
\usepackage{dcolumn}% Align table columns on decimal point
\usepackage{bm}% bold math

\begin{document}

\def\ket#1{\langle#1\mid}
\def\bra#1{\mid#1\rangle}

\title{Entanglement and quantum phase transition in the extended Hubbard model}

\author{Shi-Jian Gu$^{1,2}$}
\author{Shu-Sa Deng$^1$} \author{You-Quan Li$^2$} \author{Hai-Qing Lin$^1$}
\affiliation{$^1$Department of Physics, The Chinese University of
Hong Kong,
    Hong Kong, China}
\affiliation{$^2$Zhejiang Institute of Modern Physics, Zhejiang
University,
    Hangzhou 310027, P. R. China}

\begin{abstract}

We study quantum entanglement in one-dimensional correlated
fermionic system. Our results show, for the first time, that
entanglement can be used to identify quantum phase transitions in
fermionic systems.

\end{abstract}
\pacs{03.67.Mn, 03.65.Ud, 05.70.Jk}

\date{\today}
\maketitle

Quantum entanglement, as one of the most intriguing feature of
quantum theory, has been a subject of much studies in recent
years, mostly because its nonlocal connotation\cite{ABinstein35}
is regarded as a valuable resource in quantum communication and
information processing \cite{SeeForExample,MANielsenb}.
%For the application purpose, much attention have been focused
%on relevancy of quantum entanglement to realistic systems.
One important issue is whether there exists any relation between
quantum entanglement and quantum phase transitions \cite{Sachdev}.
Several groups investigated this problem by study quantum spin
systems
\cite{KMOConnor2001,XWang_PRA_64_012313,GLagmago2002,AOsterloh2002,TJOsbornee,GVidal2003,Shi,JVidal,Syljuasen,SJGuup}.
For example, the work of Osterloh {\it et al.}\cite{AOsterloh2002}
and Osborne and Nielsen\cite{TJOsbornee} on the spin model showed
that the entanglement of two neighboring sites displays a sharp
peak either near or at the critical point where quantum phase
transition undergoes.

On the other hand, real systems consist of moving electrons with
spin so to explore the relation between quantum entanglement and
quantum phase transition in fermionic system is necessary.
Previously, there are couple of works studied entanglement in
fermionic lattices
\cite{JSchliemann_PRB_63_085311,PZanardi_PRA_65_042101}, but they
did not discuss its relation to quantum phase transition. In this
Letter, in the framework of one-dimensional extended Hubbard
model, we study the change of symmetry in the ground state on
passing the phase boundary from the point view of quantum
entanglement, and demonstrate that entanglement is an unique
quantity to describe quantum phase transitions in this system. The
one-dimensional extended Hubbard model (EHM) is defined by the
Hamiltonian

\begin{eqnarray}\label{eq:Hamiltonian}
H=-\sum_{\sigma,j,\delta}c^\dagger_{j,\sigma}c_{j+\delta,
\sigma}+U \sum_j n_{j
 \uparrow}n_{j \downarrow}+V\sum_j n_j n_{j+1}.
\end{eqnarray}
In Eq. (\ref{eq:Hamiltonian}),
$\sigma=\uparrow,\downarrow;\,j=1,\dots, L; \delta=\pm 1$,
$c^\dagger_{j \sigma}$ and $c_{j \sigma}$ are creation and
annihilation operators of electron with spin $\sigma$ at site $j$,
respectively. $U$ and $V$ define the on-site and nearest-neighbor
Coulomb interactions.
%the model plays a crucial role for
%understanding many physical phenomena in condensed matter physics.
The EHM is a prototype model in condensed matter theory for it
exhibits a rich phase diagram \cite{Emery,Solyom,Lin} where
various quantum phase transitions occur between symmetry broken
states. These states include the charge-density-wave (CDW), the
spin-density-wave (SDW), and phase separation (PS). By calculating
the entanglement as functions of electron-electron interaction $U$
and $V$ as well as fermion concentration $N/L$, we show that
quantum phase transitions can be identified at places where local
entanglement is extremum or its derivative is singular. Our
results, part of which are based on the exact solution of the
one-dimensional Hubbard model, are useful for people to explore
quantum entanglement and quantum phase transition via other
approaches for interacting many-fermion systems.

For spin-1/2 fermion system, there are four possible local states
at each site, $|\nu\rangle_j = |0\rangle_j,\,|\uparrow\rangle_j,\,
|\downarrow\rangle_j,\,|\uparrow\downarrow\rangle_j$. The
dimension of the Hilbert space for a $L$-site system is then
$4^L$, and $|\nu_1,\,\nu_2\,\cdots\nu_L\rangle=
\prod_{j=1}^L|\nu_j \rangle_j$ are its natural basis vectors.
%Therefore any state can be expressed as a superposition of these
%basis vectors.
We consider local density matrix of the ground state,
$|\Psi\rangle$, which is a reduced density matrix $\rho_j={\rm
Tr}_j|\Psi\rangle\langle\Psi|$, where ${\rm Tr}_j$ stands for
tracing over all sites except the $j$th site. Accordingly, the von
Neumann entropy $E_v$ calculated from the reduced density matrix
$\rho_j$ measures the entanglement of states on the $j$th site
with that on the remaining $L-1$ sites. It is called the local
entanglement\cite{PZanardi_PRA_65_042101} for it exhibits the
correlations between a local state and the other part of the
system. Since Hamiltonian (\ref{eq:Hamiltonian})
%conserves total spin and charge, and it
is invariant under translation, the local density matrix $\rho_j$
is site independent,
%The density matrix $\rho_j$ therefore has the following form
\begin{eqnarray}
\rho_j= z\bra{0}\ket{0} + u^+\bra{\uparrow}\ket{\uparrow}
         + u^- \bra{\downarrow}\ket{\downarrow}
          + w \bra{\uparrow\downarrow}\ket{\uparrow\downarrow} ~,
\end{eqnarray}
with
%Parameters in the density matrix $\rho_j$ represent local
%populations of four states, i.e.,
\begin{eqnarray}
w&=&\langle n_{j\uparrow}n_{j\downarrow}\rangle
 = {\rm Tr}(n_{j \uparrow}n_{j \downarrow}\rho_j), \nonumber\\
  u^+&=&\langle n_\uparrow\rangle - w,  \;\;
   u^-=\langle n_\downarrow\rangle - w, \nonumber\\
    z&=&1 - u^+ - u^- -w
    = 1 - \langle n_\uparrow\rangle - \langle n_\downarrow\rangle + w ~.
\end{eqnarray}
%where $\langle n_\uparrow\rangle$ and $\langle
%n_\downarrow\rangle$ are electron density with spin up and spin
%down respectively. Thus the key point in discussing the local
%entanglement is to study the density distribution of the four local states.
Consequently, the corresponding von Neumann entropy (or the local
entanglement which we call hereafter)
\begin{eqnarray}
E_v=-z\log_2z-u^+\log_2u^+ -u^-\log_2u^- -w\log_2w \nonumber
\end{eqnarray}
Clearly, the local entanglement combines four quantities which are
all important to decide the physical properties of the system. We
discover, to be shownn below, that this simple expression plays
more general and important role for the understanding of the
system than other single parameter.

%%%%%%%%%%%%%%%%%%%%%%%%%%%%%%%%%%%%%%%%%%%%%%%%%%%%%%%%%%%%%%%%%%
% Figure 1: the dependence of Ev on U and V
%%%%%%%%%%%%%%%%%%%%%%%%%%%%%%%%%%%%%%%%%%%%%%%%%%%%%%%%%%%%%%%%%%
\begin{figure}
\includegraphics[width=8.5cm]{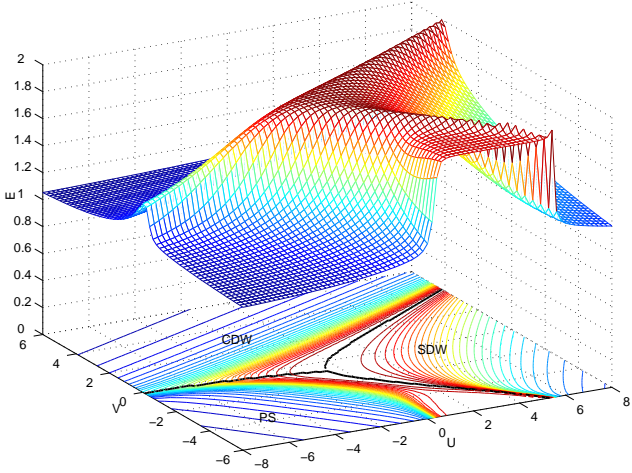}
\caption{\label{figure_uvent} The changes of symmetry in the
ground state wavefunction is analyzed by considering the quantum
correlation between local site and other part of the system. The
curved surface denotes $E_v$'s dependence on $U$ and $V$, and
colored curves on $E_v=0$ plane constitutes a contour map. Three
solid lines on the plane denote the the local extremum of a
transact of ``mountain" surface. Clearly, three main symmetry
broken phases (CDW, SDW and PS) can be sketched out from the
contour map. Superconducting phase could not be identified, due to
the fact that the broken symmetry is associated with off-diagonal
long-range-order. Finite size scaling analysis should be carried
out.}
\end{figure}

We start with general behavior of the local entanglement for the
half-filling ($N=L$) case.
%In order to study its relation to the quantum phase transition,
%we map the surface configuration of $E_v$ into contour lines.
In Fig. 1, we plot $E_v$ on the $U-V$ plane with its contour map.
It is remarkable to see that the skeleton of the EHM's phase
diagram \cite{Lin} can be directly obtained from the contour map.
This is by no means trivial. In conventional approach to obtain
the phase diagram of the EHM, one has to study behaviors of
different order parameter in different regions, either by
comparing ground state energy or critical exponent of correlation
function associated with broken symmetry. Whereas here, using a
single quantity, $E_v$, the global picture of the system could be
observed. Just like the proverb says, ``a drop of water reflects
the rays of the sun''. Obviously, this is not a coincident.
Rather, it reflects the underlying correlation between
entanglement and quantum phase transition behind the superposition
principle of quantum mechanics.

In order to clarify physical pictures further, we present our
studies in details at some special transects. Firstly, we study
the Hubbard model, i.e., $V=0$ in the EHM. Since the
one-dimensional Hubbard model can be solved analytically for both
finite and infinite lattices by the Bethe-ansatz method
\cite{EHLieb68}, we can study the analyticity of the phase
transition as well as checking the validity of the numerical exact
diagonalization technique used for the EHM on finite lattices.

%%%%%%%%%%%%%%%%%%%%%%%%%%%%%%%%%%%%%%%%%%%%%%%%%%%%%%%%%%%%%%%%%%
% Figure 2: Density of doubly occupied sites and local entanglement
%%%%%%%%%%%%%%%%%%%%%%%%%%%%%%%%%%%%%%%%%%%%%%%%%%%%%%%%%%%%%%%%%%
\begin{figure}
\includegraphics[width=7cm]{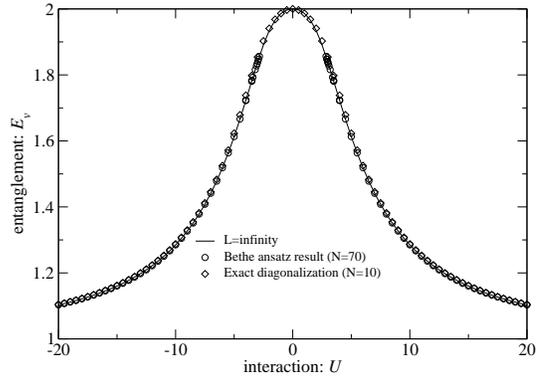}
\caption{\label{figure_ent} Local entanglement $E_v$ of the
Hubbard model at half-filling versus the on-site coupling $U$ for
different size lattice.}
\end{figure}

For the Hubbard model it is well known that its ground state is a
spin singlet and $\langle n_\uparrow\rangle=\langle
n_\downarrow\rangle=1/2$. Thus, $u^+$ and $u^-$ are both equal to
$1/2-w$ and the local entanglement is
\begin{eqnarray}
E_v=-2w\log_2w-2\left(1/2-w\right) \log_2\left(1/2-w\right).
\label{eq:localevhalffilling}
\end{eqnarray}
By making the use of particle-hole symmetry of the model, one
easily finds that $w(-U)= 1/2-w(U)$, so the local entanglement is
an even function of $U$, i.e. $E_v(-U)=E_v(U)$. In the large $U$
limit, $|U| \rightarrow\infty$, either all sites are singly
occupied ($U > 0$) so $w=0$, or half of the total sites are doubly
occupied while the other half are empty so $w=1/2$, one gets
$E_v(|U|=\infty)=1$. For finite $|U|$, hopping process enhances
$E_v$, which reaches its maximum value 2 at $U=0$ from both sides.
We plot the local entanglement $E_v$ as functions of $U$ in Fig.
\ref{figure_ent}, obtained from the Bethe ansatz method (for
$L=\infty$ and $L=70$) and exact diagonalization technique
($L=10$). The excellent agreement justifies the validity of using
small clusters for other calculations. The ground state of the
one-dimensional Hubbard model at half-filling is metallic for
$U\leq 0$, and insulating for $U>0$, so $U=0$ is a critical point
which separates metallic and insulating phases. Thus, our result
shows that the local entanglement reaches its extremum at the
critical point where the system possesses maximum SO(4) symmetry
and undergoes quantum phase transition.

Moreover, based on the Bethe ansatz solution, we can also study
the asymptotic behavior of the entanglement analytically. In the
large $U\gg 1$ region, to the third order in
$1/U^2$\cite{MTakahashib}, we have $w=4\ln
2/U^2-27\zeta(3)/U^4+375\zeta(5)/U^6$ where $\zeta$ stands for the
Riemann zeta function. Therefore the local entanglement yields the
following asymptotic behavior $ E_v=1+16\ln U/U^2+\cdots.$ Whereas
in the week coupling region $0<U\ll 1$, the density of double
occupancy becomes
$w=1/4-7\zeta(3)U/8\pi^3-93\zeta(5)U^3/2^9\pi^5$, which is
obtained by making the use of energy expansion with respect to $U$
\cite{ENEconomou79,WMetzner89}. Thus, the local entanglement near
the critical point is, $ E_v=2-\frac{1}{\ln
2}\left[\frac{7\zeta(3) U}{2\pi^3}\right]^2+\cdots.$ Clearly,
$E_v$ is analytic in the neighborhood of the critical point $U=0$.
This behavior is different from other models
\cite{AOsterloh2002,TJOsbornee,GVidal2003}.

%%------------
% figures related to filling factors
%%------------
\begin{figure}
\includegraphics[width=7cm]{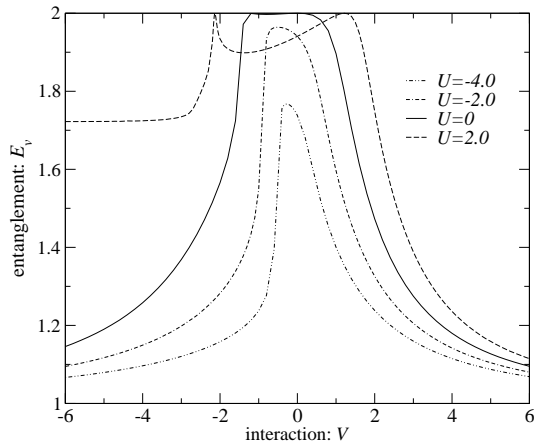}
\caption{\label{figure_lentv} Local entanglement $E_v$ versus $V$
for some values of $U$.}
\end{figure}
\vskip 1.0truecm
\begin{figure}
\includegraphics[width=7cm]{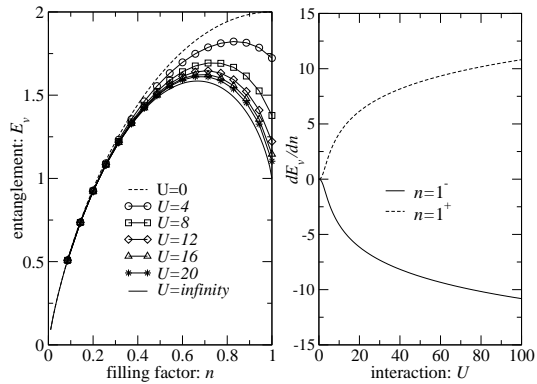}
\caption{\label{figure_lentn} Local entanglement $E_v$ as a
function of fermion concentration for the Hubbard model with
different on-site $U$(left), and the first derivative of $E_v$
with respect to $n$ at $n=1^\pm$ as  a function of $U$.}
\end{figure}

Secondly, we study the EHM at some fixed values of on-site Coulomb
interaction $U$ by varying nearest-neighbor interaction $V$. For
negative $U$, each site tends to be doubly occupied. When $V\gg
|t|$, CDW state is favored, while for $V \ll -|t|$, phase
separation occurs. Both CDW state and PS state lead to $E_v=1$.
Only in the region where $|V| \sim 0$, electron itinerant motion
dominates, tends to uniform density distribution so a local
maximum of $E_v$ occurs around $V=0$, as shown in Fig.
\ref{figure_lentv}.

For positive $U$, the physics becomes more interesting since the
model in this region is more relevant to the real materials. When
$V>0$, the competition between CDW and SDW will lead $E_v$ to an
extremum where the phase transition undergoes, due to the fact
that the local entanglement itself combines CDW order parameter
and SDW order parameter at the same time. As shown in Fig.
\ref{figure_ent} and \ref{figure_lentv}, the transition happens
along a line $U \approx 2V$, consistent with other studies of the
EHM \cite{Emery,Solyom,Lin}. When $V<0$, the formation of electron
clustering, i.e., the phase separated configuration, challenges
the SDW state. In the large $U$ and $|V|$ limit, it can be easily
shown that phase transition happens at $U=-2V$
\cite{Emery,Solyom,Lin}.
>From Fig. \ref{figure_lentv}, we also observe that there exists
singular behavior for the local entanglement at the transition
points. Moreover, for $U>0$, we find that the local entanglement
at two boundary lines is very close to 2, indicates that each of
the four local modes has nearly equal population at critical
point.

Thirdly, we study the variation of local entanglement as function
of chemical potential
%{\it The variation of the local entanglement caused by chemical
%potential}: Finally, we end up out work
by adding the term $- \mu \sum_i n_i$ to the Hubbard model.
Consequently, the total particle number of the ground state, hence
the filling factor, could be tuned. We show the relations between
local entanglement and the filling factor $n$ for various on-site
coupling $U$ in Fig. \ref{figure_lentn}. We only need to plot the
part of $n=N/L<1$ because the other part, $n>1$, could simply be
obtained by the mirror image relation, as easily seen by the
particle-hole transformation, namely,
\begin{eqnarray}\label{eq:kgjdflgdg}
E_v(n)=E_v(2-n).
\end{eqnarray}
Fig. \ref{figure_lentn} manifests that the ground state of the
half-filled band is not maximally entangled as long as $U>0$,
whereas, the maximum of $E_v$ lies in between $n=2/3$ and $n=1$.
Let us take $U=\infty$ for example. When $U=\infty$, there is no
double occupied site, which implies that $w=0$ and $u^+=u^-=N/2L$.
Hence we have an analytical expression of the local entanglement
$E_v=-(1-n)\log_2(1-n)-n\log_2(n/2)$ which has a maximum at
$n=2/3$. It is worthwhile to point out that at $1/3$ filling
(i.e., $n=2/3$) when $U=\infty$, the ground state is a singlet of
SU$(2|1)$ Lie supersymmetry algebra which possesses the maximal
symmetry allowed, while at 1/2 filling, it is a SU$(2)$ singlet.
For $U=0$ the ground state is invariant under SO$(4)$ rotation at
1/2 filling. This demonstrates that the local entanglement reaches
a maximum value at the state with maximal symmetry. Accordingly,
the maximum position for $0<U<\infty$ is expected to lie between
$n=2/3$ and $n=1$, which is numerically confirmed in Fig.
\ref{figure_lentn}.

Except at half-filling where it becomes a Mott-insulator, the
system is an ideal conductor \cite{EHLieb68}. Consequently, the
local entanglement $E_v$ is not smoothly continuous at $n=1$ for
$U\neq 0$. It is then instructive to observe the derivative of
$E_v$ with respect to $U$,
\begin{eqnarray}
\left.\frac{dE_v}{dn}\right|_{n=1^-} = -\left.(\log_2u^+-\log_2z)
    \left[\frac{1}{2}+2\frac{d\Delta E}{dU}\right]\right|_{n=1} ~,
\end{eqnarray}
where $\Delta E$ is the gap of charge excitation.
Eq.(\ref{eq:kgjdflgdg}) gives rise to
$\left.dE_v/dn\right|_{n=1^+}=-\left.dE_v/dn\right|_{n=1^-}$.
Obviously, there exists a jump in the derivative of $E_v$ across
the point of insulating phase (see Fig. \ref{figure_lentn} right)
unless $U=0$.

>From the above investigations, we find that the local entanglement
manifests distinct features at the point where quantum phase
transition undergoes. Since the local entanglement represents the
symmetry of the system to a certain extent, naturally one expects
that the maximum point of the local entanglement corresponds to
the quantum phase transition point. In the light of this
conclusion, we speculate that the maximum point in Fig.
\ref{figure_lentn} not only denotes the maximum symmetry, but
could also be a critical point separating two different phases. On
the other hand, the discontinuity properties of the local
entanglement obviously indicates a phase transition. For example,
the derivative of $E_v$ in the region of $U>0$ and $V<0$ at
half-filling of the EHM, and of the Hubbard model caused by the
shifting of the chemical potential are both not smoothly
continuous at the quantum phase transition points. This is similar
to other studies, e.g., the one-dimensional XY model in a
transverse magnetic field
\cite{AOsterloh2002,TJOsbornee,GVidal2003}, where the derivative
of the pairwise concurrence $C$ with respect to the dimensionless
coupling constant develops a cusp at the quantum phase transition
point. However, such discontinuity is not universal, as shown by
our results.

It was indicated \cite{GSTian} recently that two mechanisms may
bring about quantum phase transitions in one-dimensional
correlated fermionic systems. One is caused by the level crossing
of the ground state and the other arises from the level crossing
of the low-lying excited states where no level crossing occurs at
the ground state. In the later case, if the ground state
wavefunction is smoothly continuous with respect to the variation
of parameters that drive the quantum phase transition, the
entanglement should also be smoothly continuous at the quantum
phase transition point. On the other hand, the singularity of
wavefunction may leads to the singularity of the local
entanglement. For the former case, the level crossing of the
ground state will clearly cause the entanglement to be none
smoothly continuous at the transition point. Therefore the
continuity property of the local entanglement might be an
ancillary tool to judge the mechanism of quantum phase transition
proposed in \cite{GSTian}.

In summary, we have extensively studied the local entanglement in
the one-dimensional extended Hubbard model, characterized by the
on-site Coulomb interaction $U$, the nearest-neighbor Coulomb
interaction $V$, and band filling $N/L$. At half filling, we
calculated local entanglement as functions of $U$ and $V$. Our
results indicated that the local entanglement either reaches the
maximum value or shows singularity (or both) at the critical point
where quantum phase transition undergoes. For the traditional
Hubbard model ($V=0$), the scaling behavior close to the critical
point $U=0$ was given as manifests that the local entanglement is
an analytical function of $U$. The asymptotic behavior of the
local entanglement at the strong coupling limit,
$U\rightarrow\infty$, was also given. Furthermore, we analyzed the
local entanglement by varying nearest-neighbor $V$ while keep $U$
fixed, and found that the local entanglement is not smoothly
continuous in some critical regions. Finally, we studied the
dependence of the local entanglement on the filling factor for the
Hubbard model. The variation of the local entanglement caused by
the shifting of the chemical potential showed that the local
entanglement reaches maximum at filling factor $n$ between 2/3 and
1. For any finite $U$, the local entanglement develops a cusp at
$n=1$. In the strong coupling limit, the 1/3 filled band has the
maximum local entanglement, suggesting that the ground state with
maximal symmetry possesses the maximum magnitude of the local
entanglement.

This work is supported by the Earmarked Grant for Research from
the Research Grants Council (RGC) of the HKSAR, China (Project
CUHK 4246/01P \& 4037/02P). YQL is supported by the NSF China No.
10225419 \& 90103022.


\begin{references}

%EPR
\bibitem{ABinstein35}
A. Einstein, B. Podolsky, and N. Rosen, Phys. Rev. {\bf 47}, 777
(1935).


%Entanglement, Teleportation, Quantum Information
\bibitem{SeeForExample}
See review article by C. H. Bennett and D. P. Divincenzo, Nature
{\bf 404}, 247 (2000).

\bibitem{MANielsenb}
M. A. Nielsen and I. L. Chuang, {\it Quantum Computation and
Quantum Information} (Cambridge University Press, Cambridge,
2000).


%Quantum Phase Transition
\bibitem{Sachdev}
S. Sachdev, {\it Quantum Phase Transitions}, (Cambridge University
Press, Cambridge, UK, 2000).


%Entangled Rings(spin, spinless, etc.)
\bibitem{KMOConnor2001}
K. M. O'Connor and W. K. Wootters, Phys. Rev. A {\bf 63}, 052302
(2001).

\bibitem{XWang_PRA_64_012313}
X. Wang, Phys. Rev. A {\bf 64}, 012313 (2001).


%Quantum Phase Transition
\bibitem{GLagmago2002}
G. Lagmago Kamta and Anthony F. Starace, Phys. Rev. Lett. {\bf
88}, 107901 (2002).

\bibitem{AOsterloh2002}
A. Osterloh, Luigi Amico, G. Falci and Rosario Fazio, Nature {\bf
416}, 608 (2002).

\bibitem{TJOsbornee}
T. J. Osborne and M. A. Nielsen, Phys. Rev. A {\bf 66},
032110(2002); quant-ph/0109024 (2002).

\bibitem{GVidal2003}
G. Vidal, J. I. Latorre, E. Rico, and A. Kitaev, Phys. Rev. Lett.
{\bf 90}, 227902 (2003); J. I. Latorre, E. Rico, and G. Vidal,
quant-ph/0304098 (2003).

\bibitem{Shi}
Yu Shi, Phys. Lett. A {\bf 309}, 254 (2003).

\bibitem{JVidal}
J. Vidal, G. Palacios, and R. Mosseri, cond-mat/0305573 (2003).

%XXZ model
\bibitem{SJGuup}
S. J. Gu, H. Q. Lin and Y. Q. Li, Phys. Rev. A {\bf 68}, 042330
(2003).

\bibitem{Syljuasen}
O. F. Sylju{\aa}sen, Phys. Rev. A {\bf 68}, 060301 (2003).


%Fermion model
\bibitem{JSchliemann_PRB_63_085311}
J. Schliemann, D. Loss, and A. H. MacDonald, Phys. Rev. B {\bf
63}, 085311 (2001). J. Schliemann, J. Ignacio Cirac, M. Ku\'s, M.
Lewenstein, and D. Loss, Phys. Rev. A {\bf 64}, 022303 (2001).

\bibitem{PZanardi_PRA_65_042101}
P. Zanardi, Phys. Rev. A {\bf 65}, 042101 (2002);
%\bibitem{PZanardi_JPA_35_7947}
P. Zanardi and X. Wang, J. Phys. A: Math. Gen. {\bf 35}, 7947
(2002).


%Extended Hubbard Model
\bibitem{Emery}
V. J. Emery, pp. 247-303 in {\it Highly Conducting One-Dimensional
Solids}, edited by J. T. Devreese {\it et al.} (Plenum, New York,
1979).

\bibitem{Solyom}
J. Solyom,  Adv. in Phys. {\bf 28}, 201 (1979).

\bibitem{Lin}
H. Q. Lin, Gagliano, D. K. Campbell, E. H. Fradkin, and J. E.
Gubernatis, in {\it The Hubbard Model: Its Physics and
Mathematical Physics}, edited by D. Baeriswyl {\it et al.}, pp.
315-327; see also, H. Q. Lin, D. K. Campbell, and R. T. Clay,
Chin. J. Phys. {\bf 38}, 1(2000).

\bibitem{EHLieb68}
E. H. Lieb and F. Y. Wu, Phys. Rev. Lett. {\bf 20}, 1445 (1968).

\bibitem{MTakahashib}
M. Takahashi, {\it Thermodynamics of one-dimensional Solvable
Models} (Cambridge University Press, Cambridge, 1999).

\bibitem{ENEconomou79}
E. N. Economou and P. N. Poulopoulos, Phys. Rev. B {\bf 20}, 4756
(1979).

\bibitem{WMetzner89}
Walter Metzner and Dieter Vollhardt, Phys. Rev. B {\bf 39}, 4462
(1989).


%Quantum phase transition and excited state level crossing
\bibitem{GSTian}
G. S. Tian and H. Q. Lin, Phys. Rev. B {\bf 67}, 245105 (2003).

\end{references}
\end{document}